# Femoral Neck Angle Impacts Hip Disorder and Surgical Intervention: A Patient-Specific 3D Printed Analysis

Katie McFarlane
College of Maritime Studies and Management,
Chiang Mai University,
Samut Sakhon, Thailand
+66-6-575-89777
katie@wavertree.org[†]

Thanapong Chaichana
College of Maritime Studies and Management,
Chiang Mai University,
Samut Sakhon, Thailand
+66-6-567-28999
thanapong@wavertree.org*

Zhonghua Sun
Discipline of Medical Radiation Sciences, School of Molecular and Life Sciences, Curtin University,
Perth, Western Australia, Australia
+61-8-9266-7509
z.sun@curtin.edu.au

Joseph Neil Dentith
Department of Architecture and Built Environment, Northumbria University,
England, United Kingdom
+44-191-227-3856
joseph.dentith@northumbria.ac.uk

Philip Brown
Royal Liverpool University Hospital,
England, United Kingdom
+44-151-760-2000
pbrown25@nhs.net

## ABSTRACT
The purpose of this study is to investigate the femoral neck angulation for prediction of the complication associated with dynamic hip screw (DHS) surgery and hip deformity. Three sample patients' MRI images were selected to calculate the femoral neck angles. A total of six femur head geometries were reconstructed and three dimensional (3D) models printed. The calculation of neck angles was done in both computer models and 3D-printed models. Our results showed that 3D-printed models achieved high accuracy and provided the physical measurements, when compared to the computer models could not confirm. Neck angulations related to uncomplicated DHS surgery ranged between 129°-139°, and non-deformity of normal neck angles ranged between 120°-135°. Our study indicated that patient-specific 3D printed femoral head models provide useful information for medical education and assist DHS surgery. Further research based on a large sample size is necessary.

• **Computing methodologies** → **Computer graphics** → **Rendering** → **Visibility**.

*Corresponding author: thanapong.c@cmu.ac.th

## 1. INTRODUCTION
Digital imaging data increasingly involved in computer-based methods and 3D printing in medicine for disease diagnosis [1],[2]. Patient specific 3D printed femur head models provide important anatomical structure for hip fracture fixation [3]. The personalised 3D printed models will benefit for orthopaedic trainees in regard to increase valuable information prior to perform the dynamic hip screw (DHS) surgery [4]. Femoral neck angulation is acceptable factor to determine femur conditions. Figure 1 shows a realistic 3D printed femur model of our very own modelling created from patient 3 digital imaging data and example of DHS technique.

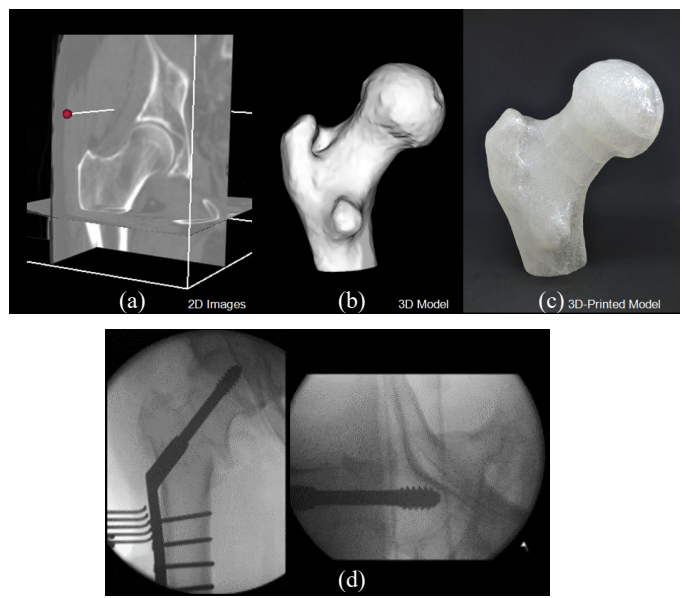

**Figure 1. Computer-aided surgery: (a) 2D medical imaging, (b) 3D surface, (c) 3D-printed model, (d) DHS approach.**

DHS is a certain type of orthopaedic implantation and fixation for personalised hip fracture. It can dynamically move the femur head components used for internal fixation of femur fractures [3]. DHS is a common technique for hip surgery in elderly patients.

McDonald et al. [5] discovered 3D-printed model and 3D printing technology in physiotherapy, the opportunities and challenges in medical education and assistive technology. They produced a 3D-printed model and used it as assistive device in clinical setting for physical therapists to treat their patients. Their works contributed in knowledge toward understandings in the use of 3D-printed geometries with practical applications in clinical settings. Those successfully assisted a personalised-crutch hand grip for training device. They concluded that 3D-printed models brought a huge benefit into medicine and improvement of healthcare service.

Chepelev et al. [6] proposed a guideline for medical 3D printing approaches. They performed a review of medical 3D printing literature and standard meeting prior to result their guidance. The recommendation reports major concerns included imaging data and segmentation, 3D model, 3D-printed model and the usage in healthcare service. Hence, the present guidance was clearly in-line with our work, as shown in Figure 1.

Poole et al. [7] explored a hip fracture is importantly caused by central osteoporosis defects in femoral head bone. They used 3D imaging computer approach to analyse different patterns of bone mapping region of interest to determine the fracture. Their study data populated more than half upon female femur head models. Statistical information of cortical bone and trabecular bone was supported their analysed results. Poole and his team concluded that their findings supported hypothesis of femur fracture and risk related to location of focal osteoporotic defects.

Audigé et al. [8] introduced radiographic imaging approach to quantify the movement of DHS surgery. They assumed that DHS change can count upon the femoral flexion and rotation using radiographic images positioned in anteroposterior (AP) setup. Mathematical calculation of DHS changed in XY coordinates of AP view was used to compute the DHS locations adjusted in millimetre. They verified their quantified results with synthetic femur bone fixed with DHS approach. They concluded that DHS relocation can calculate with using the proposed approach.

Kun [9] studied a fused deposition modelling (FDM) technology for 3D printing. He assumed that FDM technology will benefit to reverse engineering for 3D printer reproduced an accurate 3D model. Also, this may be advantage to medical technology, engineering and medicine. FDM is a 3D printing process, used a continuous filament of a thermoplastic material [9], and it is a current 3D printing technology. He concluded that after learning a FDM technology was given himself an opportunity to develop his own 3D printer. However, an open source Prusa I3 3D printer has used a fused filament fabrication (FFF) technology under the RepRap project that started in England for a low-cost 3D printer from University of Bath [10].

Marco et al. [11] investigated femur fractures using computer simulation. They purposed that recent computer technology can use to assist surgery. They used computer to create different models of femur fractures. That can predict the fracture paths and also the growing lines of cracking locations. Their computational results were validated with synthetic models of femur fractures. Marco and his team concluded that computer simulation and computer technology provided the accurate results to predict human femur fractures.

Currently, the synthetic bones and cadaveric bones are still using in clinical, education and for orthopaedic trainees to study and construct a plan for hip disease and surgery. The realistic 3D-printed femur geometries of personalised data would bring new opportunities into medical education that overcomes complicated implantation procedure of hip fracture fixation.

In consequence, this work is a preliminary study of 3D printing model of femur geometries. The aim is to calculate femoral angulation using patient specific 3D-printed femur head models, which purposed to predict the impact of neck angle upon hip disorders and surgical intervention for DHS technique.

## 2. PATIENTS AND METHODS

### 2.1 Selection of Patients
Uncontrolled group of patients was used in this study. The three patients who underwent magnetic resonance imaging (MRI) scan were randomly selected to generate 3D femur head geometries using in-house software for visualisation, and segmentation was used 3D Slicer (Kitware Inc. New York, USA). We segmented and kept the actual surfaces of all femur geometries [12],[13]. The 3D geometries of femur head models were saved in stereolithography (STL) format.

### 2.2 Anatomical Study of Femur Angulation
The femur model was produced with demonstration of the femur head regions composed of fovea capitis, head, neck, greater trochanter, lesser trochanter, intertrochanteric crest and body. Figure 2 reveals the left femur head anatomy of patient 3. In addition, we also point out the commonly cracking location of femur fracture [14],[15]. This aims to make our analysis of femur neck angle impacts femur fracture and surgery.

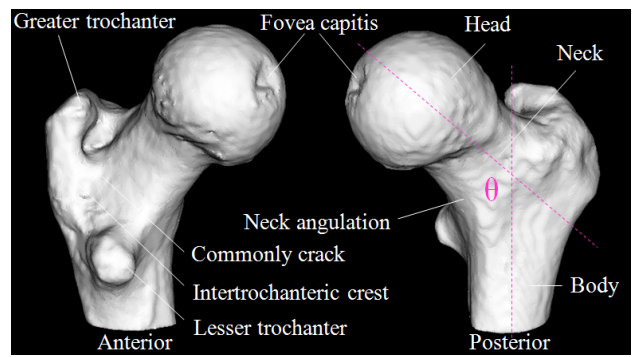

**Figure 2. Anatomical template points femur head analysis.**

### 2.3 Creating 3D Printed Femur Head Models
Home-made 3D printer was built for 3D printing of femur head models based upon the FFF technology [10]. Besides, commercial 3D printing Flashforge Creator Pro (FlashForge Corp. Zhejiang, China) was additionally used to print the femur models. The femur models were obtained in STL format, and then converted into G-code using FlashPrint 3.23 (FlashForge Corp. Zhejiang, China) and the Open-source software, Cura 15.04 (Ultimaker, Massachusetts, United States). The 3D printing slicing techniques were done in both FlashPrint and Cura. The configuration parameters were mainly followed our previous works [16], and the exception is only to configure 50°C of the Heat Beds temperature for commercial 3D printer. Figure 3 presents the 3D printing slicing approaches before taking G-code files to print with 3D printer.

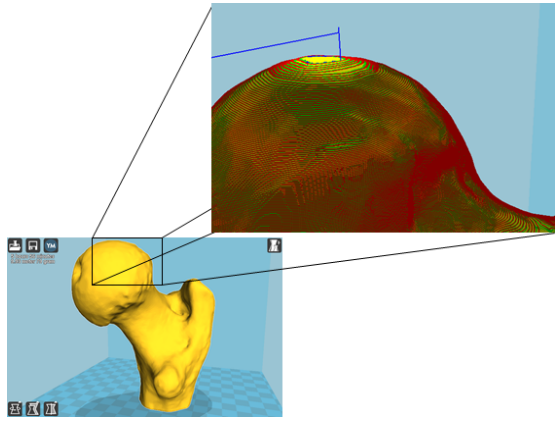

**Figure 3. Slicing technique done prior to perform 3D printing, in open-source software.**

## 2.4 Analysis of 3D Printed Femur Neck

The total of six femur head models was calculated the degree of angulation between the head and body (see at right side, centreline in Figure 2). The angulated calculations were done in both 3D surfaced-rendering models (see in Figure 1(b)) in computer, and 3D-printed models (realistic femur geometries in Figure 1(c)). The small areas of femur neck and intertrochanteric crest were identified as an impact of femur fracture (commonly crack region). Finally, the neck angulations were analysed and compared to statistical data for complication of femur DHS surgery [17].

## 3. RESULTS

Six femur head models were successfully created for 3D printing and analysing neck angulations. Figure 4 reveals the 3D-printed models of all patients. Those were analysed in this study. The femur angulations of both computer models and 3D-printed models were calculated and presented in Table 1.

We found out the 3D-printed femur models were provided good characteristics and more reliable than the models in the computer. Table 1 reports the calculation of femur neck angles. The angulated values of computer models in the computer were quite overestimated a degree of femur angulations (see in Table 1 at left hand column). Thus, calculations of neck angulations with 3D-printed models presented realistically more features, such as, manually adjustable calculations, trustable in analysed areas of the study and more accurate in anatomical areas.

**Table 1. Calculations of femur neck angles for both computer models and 3d-printed models**

| Patient Number | Femur Anatomical Angulation | Computer Model | 3D-Printed Model |
|---|---|---|---|
| 1 | Left | 131° | 129° |
|   | Right | 135° | 132° |
| 2 | Left | 139° | 134° |
|   | Right | 130° | 131° |
| 3 | Left | 139° | 135° |
|   | Right | 137° | 130° |

Additionally, normal conditions of femur neck angles were ranged 120°-135°. The hip deformations were defined more than 135° and less than 120° are in coxa valga and coxa vara conditions, respectively. As a result, the 3D-printed models demonstrated that our study populations were in normal conditions. Whilst, the neck angles calculated in the computer from computer models presented that patient 3 had occurred coxa valga conditions in both left and right femur. Besides, patient 2 had the same conditions with patient 3 at left femur, as shown in Table 1 in the column of computer models.

Alternatively, femur models presented in the flat panel display in computer were very difficult to calculate the neck angles, because the views of visualisations are limited to justify realistic surfaces and study areas. These factors can provide and prove with using the realistic 3D-printed femur models.

The impact of neck angulations of six femur models was classified using statistical data [17]. The cut-off angle was 135° for the normal setups of guide-wire insertion. So, if the angle increased 5° then the tip pins of the guide-wire moved 5mm, and do the same with the decrement distances. For our femur models, the results showed that dynamic hip screw surgery with these studied models mostly uncomplicated with DHS method. All femur angles ranged from 129° to 139°. As a result, variation factor of neck angle is ±4° calculated from the maximum value of deformity femur, and this can possibly use DHS normal setups.

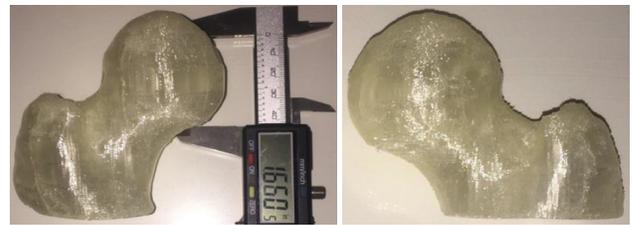

(a)                 (b)

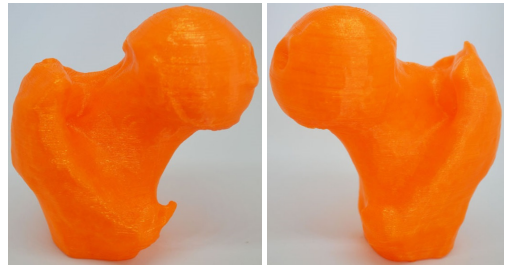

(c)                 (d)

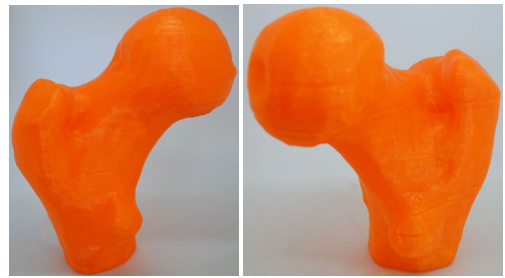

(e)                 (f)

**Figure 4. 3D-printed femur head models: from top to bottom: patient 1, 2 & 3; from column left to right: femur left & right.**

## 4. DISCUSSION

For many orthopaedic trainees, DHS for hip fracture fixation is a staple procedure [3]. Current methods for simulating the practice of DHS cannot reproduce some aspects of the procedure. The positioning of the initial guide-wire with appropriate tip-apex distance is often done by eye and with external referencing. This is not possible in a real theatre situation where image intensifiers are used. Conventional saw bone type models are not transparent and cannot reproduce the reality of the surgery. Figure 5 uncovers an actual example of the conventional saw bone model and typical image of operation anteroposterior (AP) view of a DHS in situ.

Transparent bones that are commercially available, often too expensive for training, and are made of material that is very hard and does not drill well. It is proposed that the production of a bone shape with part manufactured from a lattice or semi-transparent material may better reproduce the mechanics of the procedure. This can achieve by utilising the growing availability and accuracy of three-dimensional printing of femur models. This work serves to deliver a method that will convey a better understanding of medical modelling to improve surgical training for orthopaedic trainees.

Using an existing stereolithography file which can edit or creating a purpose built file, a low technological and low cost simulator produced to aid surgical training. A lattice used to replace the proximal 1/3 of an average size femur with gaps orientated properly and sized correctly. The appropriate angulation orthogonally should be "transparent", allowing the guide-wire to be visualized through the material (see Figure 1(a)-(d)). This can be done with using our 3D-printed femur models. In the future use of different materials and internal spacing of the model will further refine the model to increase its usability.

Creation of this new technique is to reproduce the environment of DHS at a more cost effective and realistic manner. The femur models were anatomically correct, allowing orthopaedic students to train on a model that is identical to the patients. With the training more practical and more alike to how it is in the operating theatre, valuable skills will be learned that can be transferred into the operating theatre.

In this work, we found out the measurements that a human takes for the size and angle of a femur head was more accurate than the measurements that a computer can take for the same model. These models and their measurements helped us determine whether patient's femur is angled within healthy limits or is out of range in a way that will signify that the patient is suffering from a disease. This meant that our work had a positive impact on the training needed to operate on patient hips. With a more accurate model being produced of the patient's femur, orthopaedic trainees can work and train on a model that can be sawn, and have guide-wires inserted in the same manner that the procedure will be completed in the operating theatre.

In this study, we investigated the effectiveness of three-dimensional printing to produce an accurate model of three patient's femurs. The model produced anatomically correct, with the material that the model is produced with being able to be sawn and have guide-wires inserted. These guide-wires would be visible through the "transparent" material in the same way that would occur in the operating theatre. Our results enhanced the understanding that computational technology such as three-dimensional printing can have a positive impact on the training and teaching of orthopaedic students.

In conclusion, we have performed analysis and creation of the three-dimensional femur head models from MRI data. Femur neck angles were calculated in both computer models and 3D-printed models. Six femur models in total were created and analysed the impact of neck angles upon the dynamic hip screw surgery and hip deformity. We found out that these randomly selected three patients presented an uncomplicated hip replacement and the 3D-printed models were provided great opportunities to study actual personalised femur anatomy. Further studies of large numbers of 3D-printed femur models are necessary to warrant potential risk of femur angles upon hip diseases and DHS surgery.

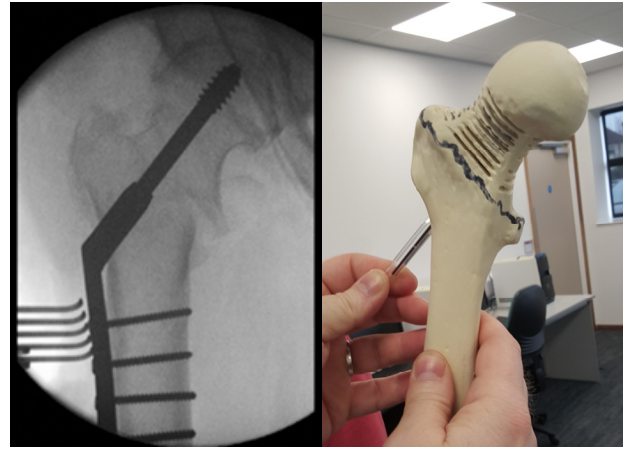

**Figure 5. Demonstrating imitation of guide-wires insertion of hip undergoing DHS fixation in AP view (left-side) used conventional saw bone model (right-side).**

## 5. ETHICAL STATEMENT
Ethics approval for this study was obtained from Liverpool Hope University's ethics committee. This research project has been approved in the category with no-human subject involved in this study. Due to the retrospective collections of de-identified MRI image data, patient consent was waived and biomedical imaging information are decently obtained to represent the digital imaging point-clouds of femoral anatomy locations used in this study.

## 6. ACKNOWLEDGMENTS
Authors gratefully acknowledge the Higher Education Innovation Fund (HEIF), Liverpool Hope University, granted funding and financial supports for 3D printing initial project. Besides, authors would sincerely appreciate assistance received from Dr Kasemsuk Sepsirisuk and Miss Anongnat Intasam to this work.